\ifx\pdfoutput\undefined
\documentclass[aps,prl,showpacs,twocolumn,superscriptaddress]{revtex4-1}
\usepackage[dvips]{graphicx}
\RequirePackage[dvipdfm,hyperindex]{hyperref}
\else
\documentclass[aps,prl,showpacs,twocolumn,superscriptaddress]{revtex4-1}
\usepackage{graphicx}
\DeclareGraphicsExtensions{.pdf,.png,.jpg,.mps}
\fi
\usepackage[english]{babel}
\usepackage{bm}
\usepackage{color}
\usepackage{amsmath}
\usepackage{natbib}
\usepackage{SIunits}
\usepackage{version}
\usepackage{longtable}

\usepackage{epstopdf}

\usepackage[english]{babel}
\usepackage{bm}
\usepackage{color}
\usepackage{amsmath}
\usepackage{natbib}
\usepackage{SIunits}
\usepackage{version}
\usepackage{longtable}

\par

\def\bea{\begin{eqnarray}}
\def\eea{\end{eqnarray}}
\def\ben{\begin{equation}}
\def\een{\end{equation}}
\def\benu{\begin{enumerate}}
\def\enu{\end{enumerate}}

\def\sss{\scriptscriptstyle\rm}

\def\br{{\bf r}}

\def\x{_{\sss X}}

\begin{document}

\title{ Fermionic correlations as metric distances:\\ A useful tool for   materials  science}

\date{\today}

\author{Simone Marocchi}
\affiliation{ Instituto de Fisica de S\~{a}o Carlos, Universidade de S\~{a}o Paulo, CP 369, 13560-970, S\~{a}o Carlos, SP, Brazil.}
\email{simonemarocchi@ifsc.usp.br}

\author{Stefano Pittalis}
\affiliation{ CNR-Istituto di Nanoscienze, Via Campi 213A, I-41125 Modena, Italy}
\email{stefano.pittalis@nano.cnr.it}

\author{Irene D'Amico}
\affiliation{ Instituto de Fisica de S\~{a}o Carlos, Universidade de S\~{a}o Paulo, CP 369, 13560-970, S\~{a}o Carlos, SP, Brazil.}
\affiliation{ Department of Physics and York Centre for Quantum Technologies, University of York, York YO10 5DD, United Kingdom}
\email{irene.damico@york.ac.uk}

\begin{abstract}
We introduce a rigorous, physically appealing, and practical way to measure distances between exchange-only correlations of interacting many-electron systems, which works regardless of their size and inhomogeneity.
We show that this distance captures fundamental physical features such as the periodicity of  atomic elements, and that it can  be used to  effectively and efficiently analyze  the performance of density functional approximations. We suggest that this metric can find useful applications in high-throughput materials design.
\end{abstract}


\maketitle

\section{I. INTRODUCTION}

The discovery of innovative materials and  engineering devices with targeted properties
involve substantial experimental and theoretical efforts.
Their progress ultimately relies on our understanding of the physics  at the nanoscale.
Atomistically, the possible constituents and their combinations are vast. One can often
focus on  the  state of electrons within the Born-Oppenheimer approximation, however
a too direct computational approach is in general unpractical,
because of the presence of many degrees of freedom and the fact that these are interrelated in a non-trivial fashion.
 Density functional theory (DFT) proposes an alternative by transforming
the problem of determining interacting many-body system properties into the
solution of the Kohn-Sham  (KS) equations, which only involve auxiliary non-interacting particles \cite{Kohn:1999,Capelle:2006,KochHolhausen:2001}.
Practically, the KS approach relies on the possibility of devising approximate forms for the exchange-correlation (xc) energy -- a functional of the particle density.
This functional embodies the effects of many-body correlations due to the intrinsic  anti-symmetry of the many-electron state and to the electrostatic
electron-electron repulsions; it also accounts for the auxiliary KS system being non-interacting.
Within this context, we wish to expose the usefulness of introducing metric spaces to analyze many-body correlations
-- when the protocol to define these spaces is both rigorous and based on quantities with a deep physical meaning.

There is an increasing interest in the use of metrics to explore quantum mechanical systems \cite{DAmico:2011,Sharp:2014,Sharp:2015,Sharp:2016, Adesso:2016, Pachos:2016, Funo:2017}, and appropriate (``natural'') metrics for particle densities, wavefunctions, and external potentials \cite{DAmico:2011,Sharp:2016}
already shed light on (previously unknown) features of the mappings at the base of the Hohenberg-Kohn theorem, the cornerstone of DFT.
Among the ultimate goals of DFT applications is the determination of properties such as total energies, ionization potentials, electron affinities,  the fundamental gaps, and lattice distances of crystalline structures. All these quantities
can  be computed accurately only if the relevant two-body correlations are properly captured by the underlying approximations.
The xc energy, at the core of the KS DFT approach, can be expressed in terms of the aforementioned two-body correlations by means of the xc-hole function as defined in the
 so-called adiabatic coupling-constant integration \cite{Kohn:1999,Capelle:2006, KochHolhausen:2001}.
Furthermore, the xc hole can be split into a correlation (c) and an exchange (x) component.
Here, we  focus on an exchange-only analysis of this quantity (more details follow below), which is useful for dealing with
relatively weakly correlated systems' ground states.
First, we  will introduce a ``natural'' distance for the x hole and show that it captures fundamental physical features such as the periodicity of  atomic elements; afterwards we will also demonstrate that it can be used to effectively and efficiently analyze the performance of density functional approximations.

\section{II. METRIC SPACE DESCRIPTION OF EXCHANGE HOLES}

Let us briefly remind the reader of a few fundamental definitions \cite{note1}. The exchange-hole (x hole) has an expression
\ben\label{nx}
n\x(\br,\br') = -  \frac{ \sum_\sigma |\gamma_\sigma(\br,\br')|^2}{n(\br)}
\een
which can be evaluated once the KS one-body reduced density matrix (1BRDM)
\ben
\gamma_\sigma(\br,\br') =  \sum_{k} f_{k \sigma} \psi_{k \sigma}(\br)\psi^{*}_{k \sigma}(\br')
\een
is known. This, in turn, only requires the knowledge of the {\em occupied} single-particle orbitals $\psi_{k,\sigma}(\br)$.
Here, $f_{k \sigma}$ are occupation
numbers and $\sigma$ is the $z$ projection of the spin index \cite{note2}.
At the denominator of Eq.~(\ref{nx}), the particle density  is  determined from
the trace $ n(\br) = \sum_\sigma  \gamma_\sigma(\br,\br) $.
Note that the calculation of the x energy, $E\x$,  can be based on the knowledge of the system-averaged  x hole, $\langle n\x \rangle$, as follows:
\bea\label{Ex}
E\x &=& 2 \pi  \int_0^{\infty}  u du~ \langle n\x \rangle(u)
\eea
where
\ben\label{sysa-h}
\langle n\x \rangle(u) := \int d\br ~ n(\br)  n\x(\br,u),\;
\een
with
\ben\label{spha-h}
n\x(\br,u)  := \frac{1}{4\pi} \int d\Omega_{ \mathbf{u} }~ n\x(\br,\br+\mathbf{u} )\;
\een
being the spherical average of the x hole, and
$\Omega_{ \mathbf{u} }$ being the solid angle defined  by $\mathbf{u}$ around $\bf r$.
Therefore, practical calculations in DFT can be enabled by providing approximations  for  $\langle n\x \rangle(u)$.
Sensible approximations must satisfy important exact conditions. In this respect, it is well known that the property
\ben\label{norm-h1}
\int_0^{\infty}  4 \pi u^2du~ \langle n\x \rangle(u) = -N\;
\een
together with the pointwise negativity condition are of outmost importance. These two properties can be combined, giving raise to the constraint
\ben\label{norm-h2}
\int_0^{\infty}  4 \pi u^2du~ |\langle n\x \rangle(u)| = N\;.
\een
{\em Crucially}, through Eq.~(\ref{norm-h2}) and by following the protocol for deriving natural metrics of Ref.~\onlinecite{Sharp:2014}, {\em these same conditions  allow us to define the natural distance} between two given system-averaged x-hole functions,
\ben\label{Dx}
D\x[ \langle  {n}^{(1)}\x \rangle, \langle  {n}^{(2)}\x \rangle] := 4 \pi
\int_0^{\infty} u^2du~  |  \langle  {n}^{(1)}\x \rangle(u) -  \langle {n}^{(2)}\x \rangle(u) |\;.
\een

Equation (\ref{Dx}) is the key result of the present work.
We emphasize that the same exact conditions that are essential to explain surprisingly good performance of even very  rough DFT approximations,
allow us to introduce a rigorous metric: we expect then this metric to capture the essential physics of exchange-only correlations.

Equation (\ref{Dx}) summarizes the difference between the exchange-only correlations of two many-body systems into a {\em single number}.
While differences of exchange energies  could be thought too as ``single numbers'' to estimate the difference between the exchange in two systems,
Eq.~(\ref{Dx}) not only rigorously satisfies the mathematical properties of a distance \cite{distance_properties} but  also
enables a comparative analysis of the systems that is far more detailed than the claim that they have the same exchange energy -- the examples
illustrated below will provide a vivid illustration of this point.
By the metrics' axioms, $D_x=0$ if and only if the two systems considered have
the {\em same} system-averaged x hole (modulo irrelevant differences over sets of vanishing measure).
For non vanishing distances,  Eq.~(\ref{Dx})  implies {\em a well-defined maximum}, given by the sum of the two systems' particle numbers.
This can be evinced from Eqs.~(\ref{Dx}) and (\ref{norm-h2}) by considering two systems of particle numbers $N_1$ and $N_2$ for which the system-averaged x holes do not overlap: in this case $D\x= N_1+N_2$. Because the system-averaged x holes have a definite sign, this also corresponds to the maximum distance between the two systems. This property implies that the x-hole distance between two systems gives us a {\em non-arbitrary} ``absolute'' measure of their closeness, as their distance
can be recast in terms of a  {\em percentage of their maximum possible distance}.

Furthermore, Eq.~(\ref{Dx}) implies a very effective geometrical structure of the physical Fock space.
Consider the application of Eq.~(\ref{Dx}) to compute the distance between the {\em exact} system-averaged x holes of two different systems.
This distance represents a measure of the difference of the exchange-only correlations between two systems.
A system with no particles may be thought of as a point, say, at the center of the Fock space.
Because of Eq.~(\ref{norm-h2}), all the other systems will be distributed at  a fixed distance
equal to the number of particles in the systems. Thus, the overall Fock space can be thought of as the union of disjoint ``onionlike'' shells:
systems with same number of particles are on the same shell; systems whose external potentials
differ only by a constant are separated by a vanishing distance (i.e., they occupy the same point) as the orbitals and therefore the 1BDM and corresponding
particle densities do not change. Exchange holes and therefore their distances are unchanged if each single-particle orbital is multiplied by the same constant phase.
This embodies the fact that both the Schr\"odinger equation and the DFT framework are invariant under global gauge transformations \cite{note3}.
Systems will be on different shells if they have different particle numbers:
the distances  acquire minimum value (i.e., the absolute value of the
difference of the shell radii) if the systems ``face each other,'' and they acquire maximum value (i.e., the sum of the shell radii) if the systems are ``on opposite poles'' \cite{note4}. Of course, the configurations which generate maximum and -- for systems on different shells -- minimum distances are not unique.

Finally, let us consider the evaluation of Eq.~(\ref{Dx}) using some approximate $ \langle  {n}\x \rangle$.
Since Eq.~(\ref{norm-h2}) must be fulfilled, proper approximations preserve the mentioned onionlike structure of the Fock space.
Also the minimum and maximum distances are unchanged, but the configurations at which these occur may vary from the exact case.
The errors due to the approximation may be viewed as fictitious displacements of the systems from their exact locations on the aforementioned shells.
Having the possibility to quantify these errors through a rigorously defined  distance
that can also be visualized is, {\em per se}, very appealing.
In the rest of this paper, we will give explicit examples of how powerful this approach can be.

\section{III. NUMERICAL RESULTS}

We start by considering a set of systems for which the exact x holes can be calculated: we will discuss the exact results as well as compare and contrast these with corresponding results from DFT approximations.  Here we shall consider popular approximations for $ \langle  {n}\x \rangle$: the local-density approximation (LDA),  the generalized gradient approximation (GGA), and the meta-GGA (MGGA).
The LDA takes as a reference the xc energy densities of the homogeneous electron gas; GGA and MGGA
are  nonempirical refinements which aim at capturing the effects of  system inhomogeneities -- those neglected within the LDA --
while progressively satisfying a larger set of exact conditions.  LDA forms  make use only of  the particle density $n(\br)$  as input;
GGAs also use the reduced dimensionless gradient, $s(\br) = |\nabla n(\br )|/\{2 \left[ 3 \pi^2 \right]^{1/3} n(\br)^{4/3}\}$; $n(\br)$ and $s(\br)$,
the kinetic-energy density $\tau = \sum_{k \sigma} f_{k \sigma}  |\nabla \psi_{k \sigma}(\br) |^2 $, and, possibly, the Laplacian of the particle density may be exploited in MGGAs. MGGA forms are then considered to be the most accurate approximations among these three.
As representative approximations for $\langle n\x \rangle$, we choose the versions of the Perdew-Wang LDA and of the Perdew-Burke-Ernzerhof GGA by Ernzerhof and Perdew \cite{Ernzerhof:1998} and the version of the Tao-Perdew-Staroverov-Scuseria MGGA by Constantin et. al~\cite{Constantin:2006}.

\begin{figure}
\includegraphics[width=\linewidth]{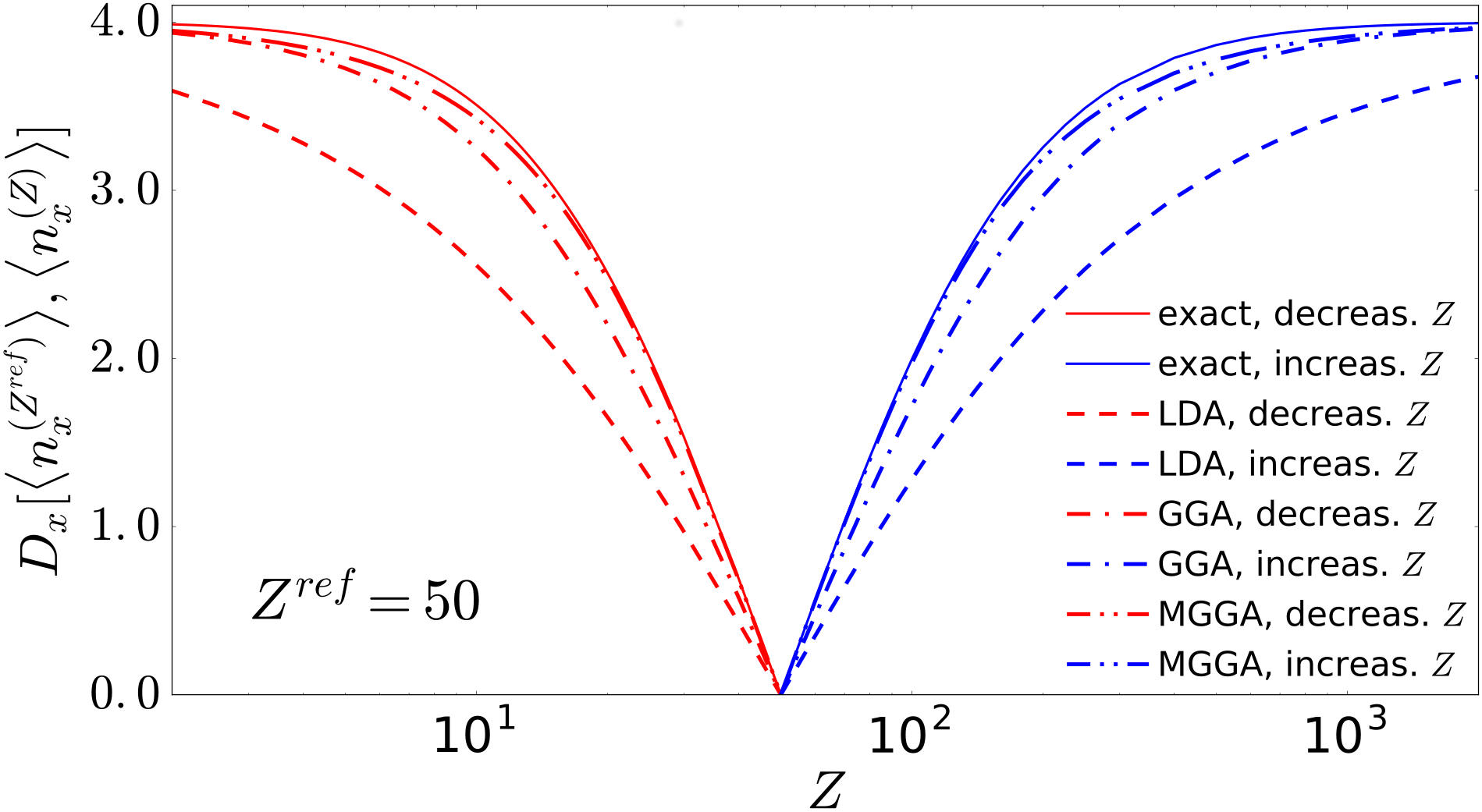}
\caption{ The x-hole distance $D_x$ from the reference state $Z^{ref}=50$ is plotted against the atomic number $Z$ for the heliumlike ion series. The exact results correspond to the solid lines, LDA to the dashed lines, GGA to the dash-dotted lines, and MGGA to the dash-double-dotted lines.
}\label{helium_ex_ex}
\end{figure}

Figure~\ref{helium_ex_ex} shows the distances of the exact $ \langle  {n}\x \rangle$ (solid line) from a reference system chosen (arbitrarily) at $Z^{ref}=50$ for the isoelectronic heliumlike sequence \cite{note5}. Distances from the reference system increase monotonically for both increasing and decreasing values of $Z$.
As the distance increases, the spatial overlap of the related system-averaged x holes decreases. The system-averaged x holes $ \langle  {n}\x (u)\rangle$ describe
the system-averaged electron depletion observed at separation $u$ from a reference electron due to the effect of electron-electron exchange, so
an increasing distance $D_x$  implies systems with an increasingly different spatial exchange pattern.
When there is no overlap between these patterns, their distance saturates at its maximum, which is $D_x^{max}=4$ for the set of systems of  Fig.~\ref{helium_ex_ex}.
Next we check how the trend for the exact exchange of heliumlike ions is reproduced by the approximations (dotted, dashed, and dash-dotted lines, as labeled in Fig.~\ref{helium_ex_ex}). While the qualitative general trend is mostly reproduced, we note that, quantitatively, the fewer exact conditions  an approximation satisfies,
the higher the inaccuracy, which in fact increases as we move from MGGA to GGA to LDA.
In particular, LDA becomes unable to reproduce, even qualitatively, the saturation to maximum distance, despite considering ion's nuclear charges as large as $Z=2000$.

\begin{figure}
\includegraphics[width=\linewidth]{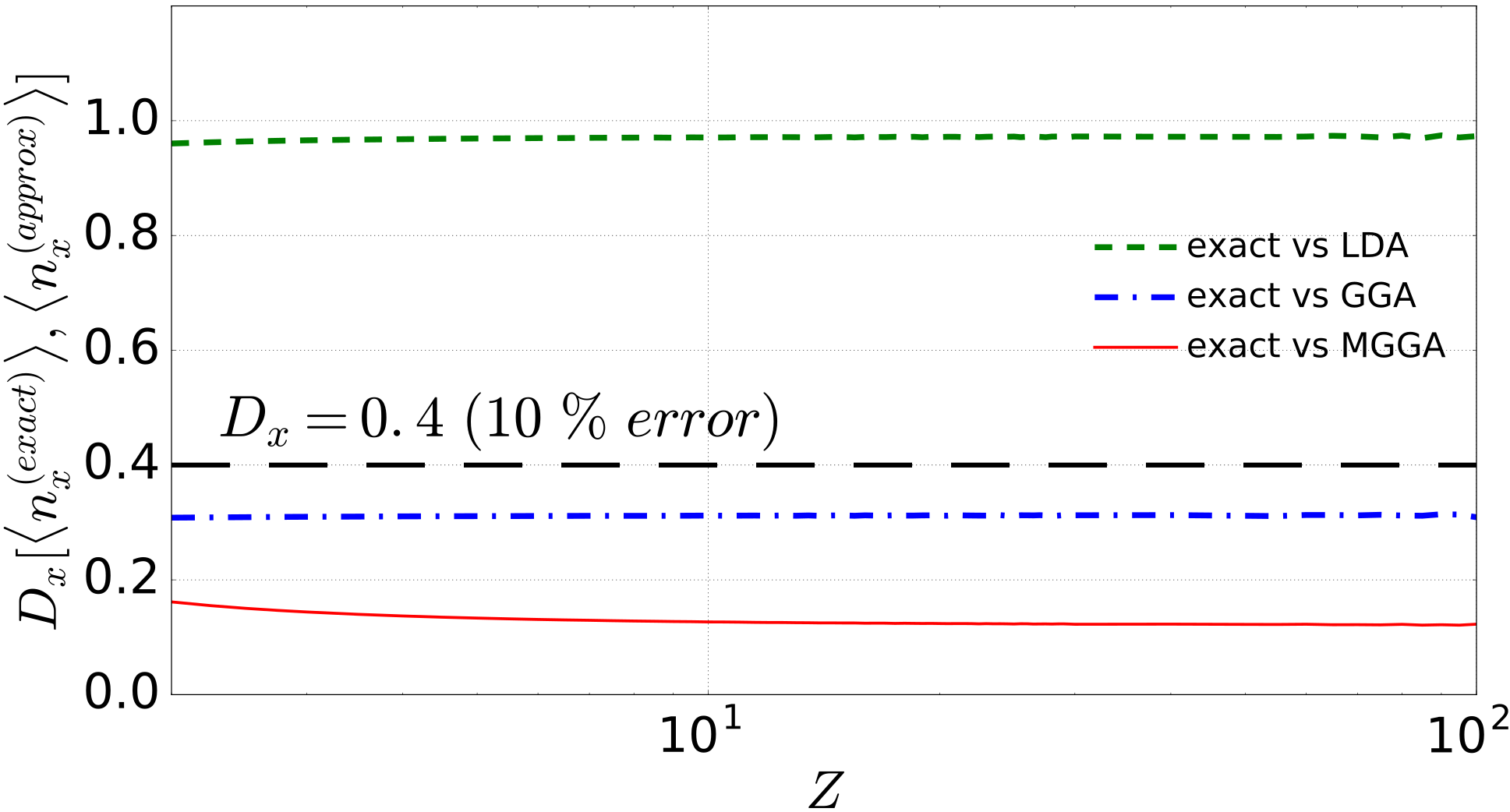}
\caption{The x-hole distance $D_x$ is plotted against the atomic number $Z$ for the heliumlike ion series. For each $Z$, the distance is calculated between the exact x hole and several approximated x holes (LDA, GGA and MGGA), as labeled. The black dashed line represents 10\% percent of the maximum distance from the exact x hole.
}\label{helium_ex_approx}
\end{figure}

Distances can also be used to perform ``point-by-point'' exact-to-approximated comparisons, by directly computing the distance between exact and approximated exchange for each system. Figure~\ref{helium_ex_approx} shows the distances of  approximated  $ \langle  {n}\x \rangle$ from the corresponding exact quantity for each ion in the isoelectronic heliumlike sequence. As the electrons get strongly confined around the
nucleus, the effect of the electron-electron interaction becomes negligible with respect to an external potential which increases linearly with $Z$.
In this way, the noninteracting limit of an infinitely charged ion is approached. Interestingly, errors with respect to the exact results quickly saturate at a finite constant.
For  LDA and GGA,  these errors may be mainly related to spurious self--interactions.
Notably, although the considered MGGA gives rather accurate x energies for two-electron systems, it is obvious that a sizable error still persists at the level of $ \langle  {n}\x \rangle$.
Importantly, the use of natural metrics allows us to {\em quantify} what we mean by ``sizable,'' by expressing the error as a {\em percentage of the maximum distance}.
In the case at hand then, a 10\% error threshold would correspond to $D_x = 0.4$ (dashed black line). We can then assert that for the heliumlike ion series, both GGA and MGGA always provide results which are closer than 10\% to the exact ones (about 7.8\% for GGA and between 4.0\% and 3.0\% for MGGA), while LDA estimates, at about 24.0\% of $D_x^{max}$, are always well above the chosen error threshold.

Consistent with the general expectation, both in Fig.~\ref{helium_ex_approx} and Fig.~\ref{helium_ex_ex}, the GGA performs in  between the MGGA and LDA; however, our method and results show in an immediate and appealing visual way how substantial is the improvement obtained in going from an LDA to a GGA. The improvement of the MGGA over the GGA is not as large as from LDA to GGA, but still significant.

For DFT practitioners, it is important to clarify under which circumstances numerically ``cheaper'' approximations could be used in place of more accurate
but computationally more involved approaches. Toward this goal,
in the rest of this paper, we show how the metric for the x hole can be used to efficiently compare the performance of different DFT approximations on
large sets of systems. In the process, we will also show how $D_x$ can be used to capture and compare physical trends within a large set of systems.
\begin{figure}
\includegraphics[width=\linewidth]{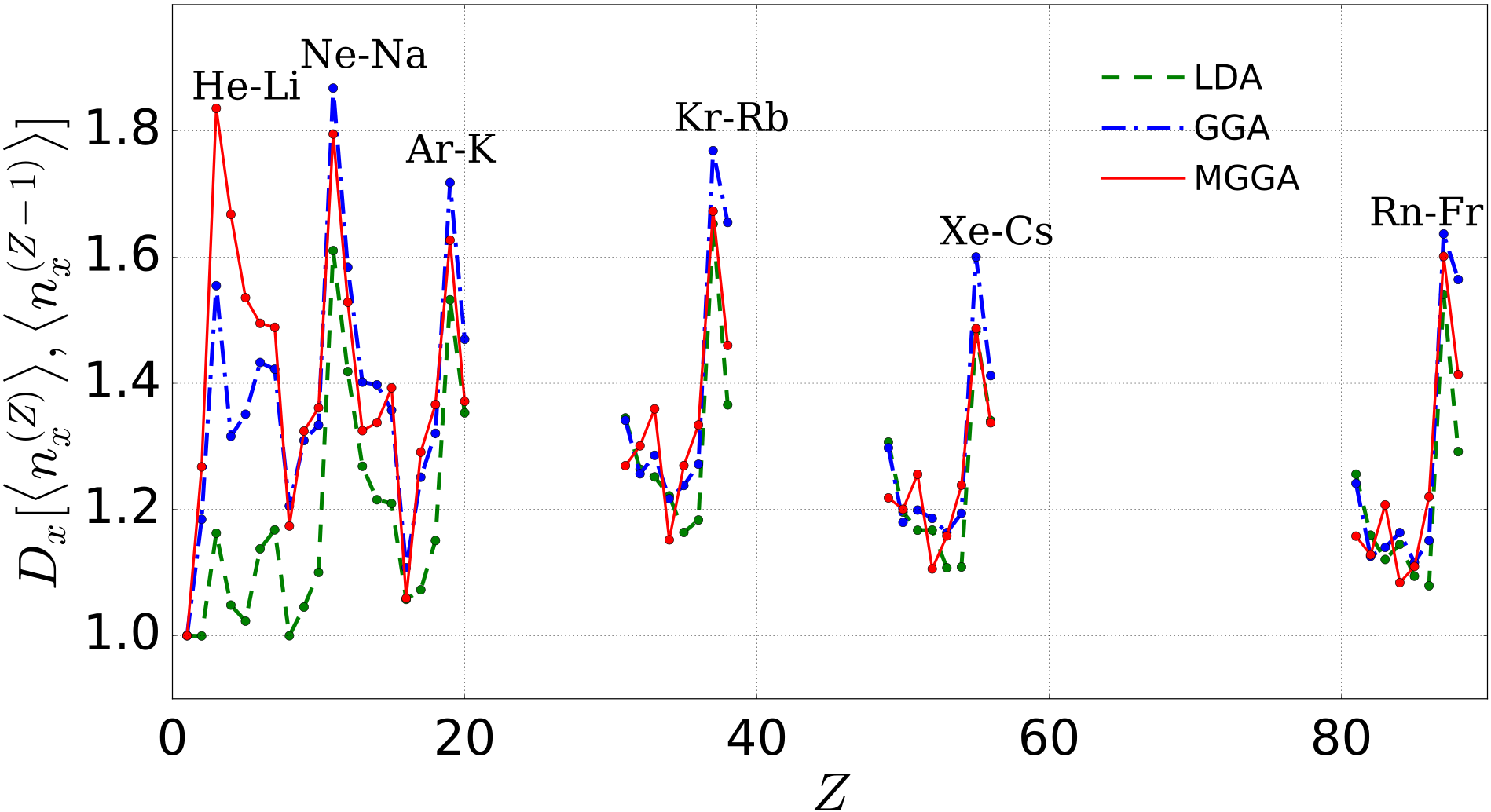}
\caption{ The x-hole distance $D_x$ between atoms with atomic numbers $Z$ and $Z-1$  is plotted against $Z$ for the $s$ and $p$ blocks of the periodic table. The distances are calculated for LDA, GGA, and MGGA, as labeled. The distances between the end and the beginning of consecutive periods are explicitly labeled with the corresponding atoms.
All the input Kohn-Sham quantities have been obtained using the APE code \cite{APE}. We allow for spin polarization by performing spin-DFT calculations~\cite{note2}.
We used logarithmic spaced grids and cubic spline interpolation \cite{Dierckx:1993} to calculate the x-hole distance between different atoms. All the densities and x-hole sum rules were tested within $10^{-4}$. }
\label{period_tab_same_approx}
\end{figure}

First we focus on physical trends within a set of systems,
and so we consider distances between x holes of {\em different} systems calculated using the {\em same} approximation.
Figure~\ref{period_tab_same_approx} shows distances between neutral atoms with atomic numbers $Z$ and $Z-1$.
Moving along the rows of the periodic table, the periodicity is well reflected in the behaviors of $D_x$ for MGGA (solid line), the most accurate approximation considered here. For example, the curves characteristically peak when considering the distance between the x holes of the last atom of one row and the first of the next (as labeled in Fig.~\ref{period_tab_same_approx}). This behavior follows from the sharp change of the corresponding atomic sizes.
The MGGA curves also display characteristic minima at every start of double occupancy in spin of the $p$ shells: as the fourth $p$ electron is introduced, the atomic radius does not change significantly. This implies that the x-hole distance from the previous atom sharply decreases.
Significant deviations are observed for LDA results for  atoms in the first two rows. We explain this by noting that self-interaction errors become larger in small systems, and electrons
of light elements tend to behave rather differently from the electrons in a homogenous gas.  GGA improves over this by accounting better for density inhomogeneity, but it is still quite poor for the smaller $Z$ values. For larger values of $Z$, the trends of LDA and GGA looks qualitatively more similar to MGGA results, although, as $Z$ increases, maximum and minimum features related to the filling of the $p$ shells get displaced with respect to MGGA positions.

Next, we wish to show how distances can lead to direct comparison between different approximations: here distances are calculated between {\em different} approximations applied to the {\em same} system, e.g., the same atom.
\begin{figure}
\includegraphics[width=\linewidth]{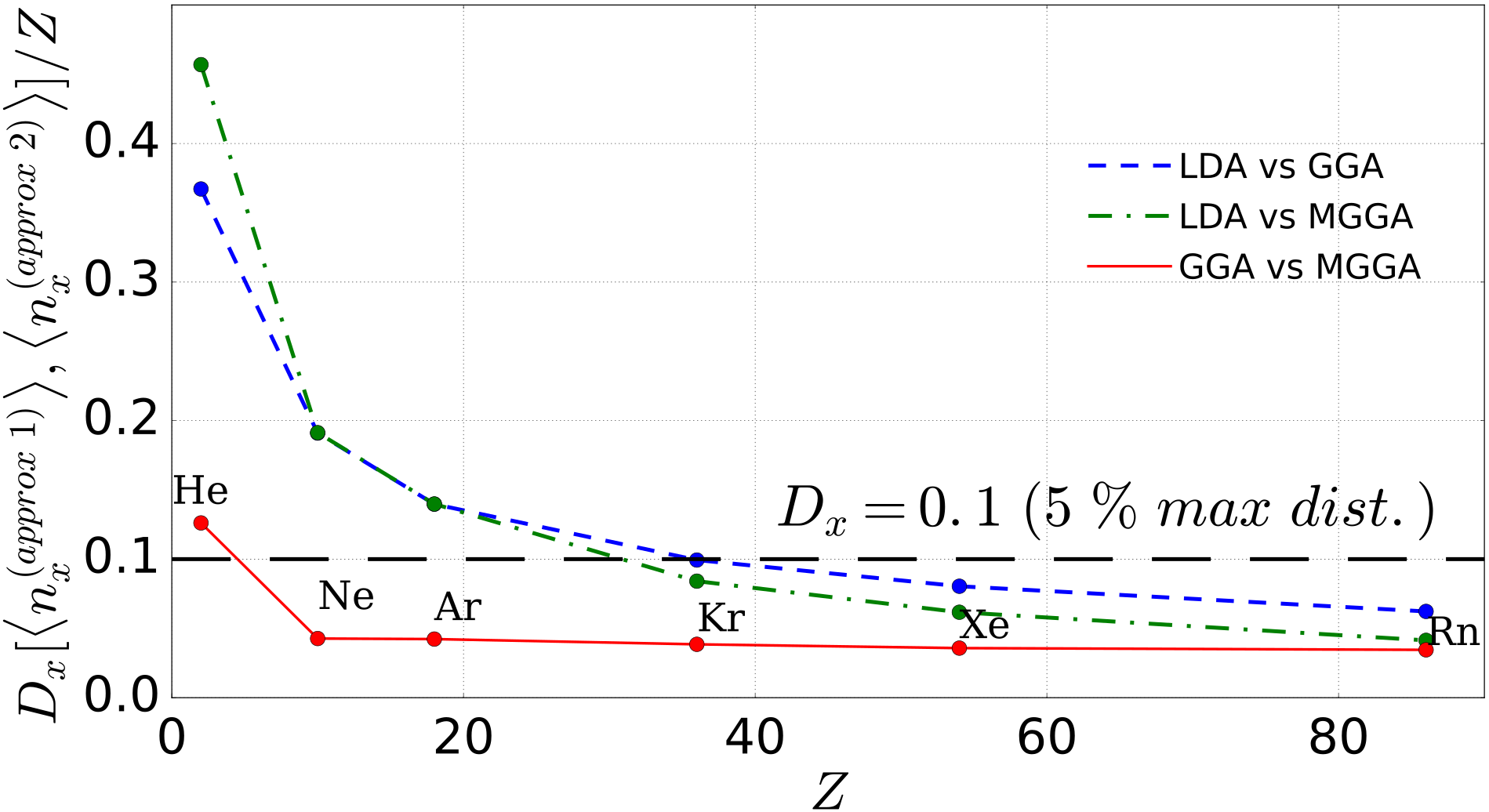}
\caption{ The x-holes distance $D_x$ is plotted  against the atomic number $Z$ for the noble gas series. The distances are calculated between different approximations to the same x hole for all atoms, as labeled, and are rescaled by the number of electrons. The black dashed line marks 5\% of the maximum possible distance.
All the input Kohn-Sham quantities have been obtained using the APE code \cite{APE}. }\label{noble_gas_approx_approx}
\end{figure}
In Fig.~\ref{noble_gas_approx_approx}, we report these distances for the noble gases. The first thing to notice is that
the distances among the various approximations decrease substantially with increasing $Z$.
This is related to the fact that in all the considered approximations, the leading contribution to the semiclassical expansion of the exchange energies
is provided through LDA \cite{Elliot09}.
The remaining differences can be attributed to high-orders contributions, more related to system inhomogeneities.
Consistently, thus, the GGA and MGGA results are closer to each other than to the LDA.
We can now define an error threshold to  establish the parameter region for which LDA and GGA would be a good-enough cheaper substitute for MGGA. As our best results are already approximated, we consider in this case a threshold of 5\% of the maximum possible distance, which corresponds in this case to $D_x/Z <0.1$ (black dashed line in Fig.~\ref{noble_gas_approx_approx}). It is immediate to see then that while LDA would be appropriate only for the heaviest three, GGA would be a good choice for all noble gases except helium.

\section{IV. SUMMARY AND CONCLUSIONS}

In summary, we have presented a way to rigorously and quantitatively compare exchange-only correlations of different systems.
We have  given evidence that by the use of a ``natural'' metrics, it is possible to effectively and efficiently characterize exchange-only correlations in many-electron systems.
Our metric based on the exchange hole could have important practical applications in evaluating DFT approximations. For example, our results suggest that among the available approximations for the system-averaged exchange-hole, the meta-GGA performs best and could be used in evaluating distances for systems widely different in size and level of inhomogeneity.
Our x-hole metric could also help guiding high-throughput materials design \cite{Curtarolo:2013},  e.g., for searching in large configurational spaces or for validating the reproducibility
of a collaborative database of electronic calculations,
independently from the different methodology, quantum
package, or hardware used \cite{Calderona:2015}. Natural metrics such as this or the one for the particle density \cite{DAmico:2011} might also be used to ensure that newly developed functionals optimize, together with the total energies, other key physical quantities, helping revert the trend recently described in \cite{Medvedev:2017}.

\section{ACKNOWLEDGMENTS}
We thank Professor Luiz Nunes de Oliveira for fruitful discussions. I.D. acknowledges support by the Royal Society through the Newton Advanced Fellowship scheme (Grant No. NA140436). I.D. and S.M. were supported by the Conselho Nacional de Desenvolvimento Cient\'ifico e Tecnol\'ogico (Grant No. 401414/2014-0) and S.P. was supported by the European Community through the FP7’s Marie-Curie International Incoming Fellowship, Grant agreement No. 623413.

\bibliographystyle{apsrev}

\begin{thebibliography}{30}
\bibitem[{not({\natexlab{a}})}]{note1}
\bibinfo{note}{For the sake of simplicity, we restrict ourselves to the cases
  for which the Kohn-Sham state is a single Slater determinant.}
\bibitem[{not({\natexlab{c}})}]{note3}
\bibinfo{note}{For discussing invariance under more general gauge
  transformations, one should admit additional couplings to proper external
  gauge fields and adopt the corresponding extensions of the DFT frameworks as
  done, for example, in current-DFT and spin-current-DFT(CDFT)~\cite{CDFT, SC1, SC2}.}
\end{thebibliography}

\end{document}